\documentclass[preprint,letterpaper,nofootinbib,superscriptaddress]{revtex4}
\usepackage{graphics}
\usepackage{epsfig}
\usepackage{amssymb}
\usepackage{mathrsfs}
\usepackage{eufrak}
\usepackage{dsfont}
\usepackage{wasysym}
\usepackage{latexsym}

\input xy
\xyoption{all}

\begin{document}

\title{The limiting behavior of the Liu-Yau quasi-local energy}
\author{P. P. Yu}
\affiliation{Department of Physics and Astronomy, Dartmouth College,
6127 Wilder Laboratory, Hanover, NH 03755 USA}
\date{\today}

\begin{abstract}
The small- and large-sphere limits of the quasi-local energy
recently proposed by Liu and Yau are carefully examined. It is shown
that in the small-sphere limit, the non-vacuum limit of the Liu-Yau
quasi-local energy approaches the expected value $\frac{4\pi}{3} r^3
{\mathfrak{T}}(e_0, e_0)$. Here, ${\mathfrak T}$ is the energy-stress
tensor of matter, $e_0 \in T_p M$ is unit time-like
and future-directed at the point $p$ located at the center of the
small sphere of radius $r$ in the limit $r \rightarrow 0$. In vacuum,
however, the limiting value of the 
Liu-Yau quasi-local energy contains the desired limit
$\frac{r^5}{90} {\mathfrak{B}}(e_0, e_0, e_0, e_0)$, where
${\mathfrak B}$ is the Bel-Robinson tensor, as well as an extra
term. In the large-sphere limit at null infinity, for isolated
gravitational sources, the Liu-Yau quasi-local energy is shown to
recover the Bondi mass and Bondi news flux, in space-times that are
asymptotically empty and flat at null infinity. The physical
validity of the Liu-Yau model in view of these results is
discussed. 
\end{abstract}



\maketitle

\section{Introduction}
\label{intro}
Among a handful of unsettled puzzles at the
foundation of Einstein's general theory of relativity, the very
basic notion of energy-momentum seems to be of everlasting interest.
Despite the triumph of the proof of the positivity of the {\it
total} gravitational energy at both spatial and null infinity
\cite{ScheonYau}, there is a lack of a well-defined notion of the
{\it local} gravitational energy-momentum density. In fact, the
equivalence principle, or, the existence of the normal co-ordinate
system, prohibits any non-trivial point-wise localizable
density. Consequently, only quasi-local quantities are
meaningful. Although various proposals have been put forward (see, for 
example, \cite{Szabados2004} for a fairly complete and up-to-date
review.), it should be noted that the study of quasi-local
quantities is still rather premature in the sense that no truly
axiomatic framework has been distilled from physics. A generally
accepted strategy of studying quasi-local quantities is to devise
such quantities and to check that they recover, in certain limiting 
situations, known properties. To be more specific, take for example
the quasi-local energy-momentum.  It has been proposed
\cite{ChristodoulouYau1988}, \cite{Szabados2004} that it satisfy the
following empirical criteria: \\

\noindent
({\bf C1}) Causality: \\
\noindent
Quasi-local energy-momentum is future-directed and non-space-like,
provided that matter, if any, satisfies the dominant energy condition
in the region enclosed by $S$.\\

\noindent
({\bf C2}) Positivity: \\
\noindent
Quasi-local energy is positive and is monotone in a suitable
sense but vanishes in flat space-times. \\

\noindent
({\bf C3}) Limiting behaviors: \\
\indent
(a) Standard sphere limit: Quasi-local energy recovers the standard
value of mass enclosed in $S$ for $S\approx S^2$ in a spherically
symmetric space-time. In particular, for a sphere centered at the
origin of Schwarzschild space-time, it coincides with the
Schwarzschild mass parameter. \\
\indent
(b) Marginally trapped surface limit:
Quasi-local mass agrees with the irreducible mass 
$\sqrt{\frac{{\rm Area} (S)}{16 \pi}}$. \\ 
\indent
(c) Small sphere limits: \\
\indent \indent
(i) in non-vacuum: on a small sphere of radius $r$ centered at any
arbitrary point $p$ in the space-time $M$, the quasi-local
energy-momentum recovers the energy-momentum of the
matter observed by an equivalence class of instantaneous observers
(the meaning of which is made precise in Sec.~\ref{vertex})
characterized by a unit time-like and future-directed $e_0 \in T_p M$, 
namely,
$\frac{4\pi}{3} r^3 {\mathfrak{T}}(\bullet, e_0)$, where
${\mathfrak{T}}$ is the energy-stress tensor. \\
\indent \indent
(ii) in vacuum: the quasi-local energy-momentum yields the
analogue of the gravitational energy-momentum observed by the class of
observers given in (i) in terms of the
Bel-Robinson tensor \cite{Bel} ${\mathfrak{B}}$, namely,
$\frac{r^5}{90} {\mathfrak{B}}(\bullet, e_0, e_0, e_0)$. \\
\indent
(d) Large sphere limits: \\
\indent \indent
(i) at spatial infinity: quasi-local energy-momentum approaches the
Arnowitt-Deser-Misner (ADM) energy-momentum in an asymptotically flat
space-like hypersurface.\\
\indent \indent
(ii) at null infinity: quasi-local energy-momentum reproduces the
standard Bondi mass $E_{\rm BS}$ and news flux
$\frac{\partial}{\partial t}E_{\rm BS}$. \\
\noindent
In fact, in the large sphere limits, quasi-local quantities are no
longer truly quasi-{\it local} as $S$ contains an infinite measure. A
perhaps more proper term here would be quasi-{\it global}.

A similar scrutiny of the quasi-local angular momentum is technically
more involved partially because the very definition of the quasi-local
angular momentum in several contexts is yet to be unanimously agreed
upon. Tentative investigations have been carried out in the past
decades with few definitive outcomes (see \cite{Szabados2004} for an
overview). The present work is thus focused on studying the
quasi-local energy, only.

Like any other construction of quasi-local energy, Liu-Yau's is
subject to reasonable reality checks in order to be a physically sound 
candidate. Among ({\bf C1})-({\bf C3}) listed above,
({\bf C1}), ({\bf C2}), ({\bf C3})-(a), (b) and (d)-(i) have been
discussed \cite{LiuYau2003}, \cite{MST2004},
\cite{LiuYauReply}. The examination of other limiting behaviors of
Liu-Yau quasi-local energy--the main body of the present work--is
presented in the following sections.

This article is structured as follows. In
Sec.~\ref{def}, a specific model of the quasi-local energy proposed
by Liu and Yau \cite{LiuYau2003} is reviewed. The small- and
large-sphere limits of the Liu-Yau quasi-local energy are closely
examined, in Sec.~\ref{small} and Sec.~\ref{large}, respectively.
Also considered, along the same lines, is the possibility of
generalizing the notion of quasi-local energy for non-isolated
gravitational sources. A summary is given in Sec.~\ref{summary}.

\section{Definition of the Liu-Yau quasi-local energy}
\label{def} The Liu-Yau quasi-local energy originated as a
continuation of Yau's mathematical work on the positivity of black
hole mass \cite{Yau2001}, \cite{ChristodoulouYau1988}. The
definition of the Liu-Yau quasi-local energy is reviewed here for
completeness.

\subsection{Physical part}
\label{defphys} Consider a closed orientable space-like 2-surface
$S$ embedded in $M$. $\forall p\in S$, $\exists$ null frame
$\{X_i\}_{i=1}^{i=n} \in {\mathds C}T_p M$ with its dual
$\{\theta^i\}_{i=1}^{i=n}\in {\mathds C}T_p^\ast M$
adapted to $S$, such that
${\mathds C}T_p M = {\mathds C}T_pS \oplus {\mathds C}T_pS^{\bot}$,
where
${\mathds C}T_p S={\rm span}_{\mathds{C}}\{X_3, X_4\}$ and
${\mathds C}T_p S^{\bot}={\rm span}_{\mathds{C}}\{X_1, X_2\}$.
(Shorthand notation
${\mathds A} E= E \otimes_{\mathds R} {\mathds A}$, where
${\mathds A}$ is a field and $E$ is a bundle, is used throughout. For
example, for ${\mathds A}={\mathds C}$, ${\mathds C}E$ is the
complexified bundle of $E$. When ${\mathds A}={\mathds R}$, however,
it is obvious that ${\mathds R} E=E$.)
The mean curvature vector of $S$ at $p$ in $M$ is then
$H=-2\mu X_1-2\rho X_2$, where $\mu=\theta^1(D_{X_3}X_4)$ and
$\rho=\theta^2(D_{X_4}X_3)$. It is worth noticing that the norm
of $H$ is, however, independent of the choice of moving frames. Indeed,
$\|H\|=\sqrt{8 \rho \mu}$, where $\rho\mu >0$ for space-like $H$.
The physical part of the quasi-local energy is then defined as
$$
E^{\rm phys}(S) \equiv \frac{1}{8\pi} \varint\limits_S \|H\| \Omega,
$$
where $\Omega$ is the volume form of $S$.

\subsection{Reference part}
\label{defref} Suppose that $S$, equipped with a Riemannian metric
$g^{\rm S}$ and Levi-Civita connection $D^{\rm S}$, has positive
sectional curvature. Then by Weyl's embedding theorem
\cite{Weyl1915}, there exists a unique isometric embedding $\iota_1:
(S, g^{\rm S}, D^{\rm S}) \hookrightarrow (M^\circ, g^\circ,
D^\circ)$, up to isometries of ${\mathds{R}}^3$, such that the
second fundamental form $II^{\circ}$ is solely determined by $g^{\rm
S}$ and is positive definite on $S$. The composition of $\iota_1$
with a successive embedding $\iota_2: {\mathds{R}}^3 \hookrightarrow
{\mathds{R}}^3_1$ gives the reference embedding $\iota \equiv
\iota_2 \circ \iota_1: S \hookrightarrow {\mathds{R}}^3_1$ of $S$
into Minkowski space-time $M^\circ$.

The same construction as in Sec.~\ref{defphys}. gives the mean
curvature vector, $H^{\circ}$, of $S$ at $p$ in ${\mathds{R}}^3_1$
whose norm is $\|H^{\circ}\|=\sqrt{8 \rho^{\circ} \mu^{\circ}}$, where
$\rho^{\circ} \mu^{\circ} >0$. Hence the
reference part of the quasi-local energy is naturally defined as
$$
E^{\rm ref}(S) \equiv \frac{1}{8\pi}
\varint\limits_S \|H^{\circ}\| \Omega.
$$

A caveat is emphasized in \cite{MST2004} and \cite{LiuYauReply}
to avoid any misleading interpretations.
Unless $S$ lies in a space-like hypersurface $\Sigma \subset M$, it
would be highly unnatural to require that $S$ be isometrically
embedded into a space-like hypersurface in $M^\circ$. That is, the
absence of this additional hypothesis may result in positive
quasi-local energy even in $M^\circ$.

One of the merits of the embedding scheme described above is that
$\iota_1$ is unique up to isometries of ${\mathds{R}}^3$ and that
$H^{\circ}$ is therefore well-defined. However, it is in general
non-trivial to obtain a complete solution to the full set of
integrability conditions for the sequence of embeddings of $S$ in
$M^\circ$. An alternative approach is to consider the co-dimension 2
embedding $\iota^\circ: S \hookrightarrow M^\circ$ at a possible
expense of uniqueness (up to isometries of ${\mathds{R}}^3_1$) unless
extra restrictions are imposed. The existence of $\iota^\circ$ is,
nonetheless, guaranteed for any conformally flat $S$, as shown in
\cite{Brinkmann1922&1923}. In Sec.~\ref{refembeddingSmall}, embeddings
of this kind are realized as the null-cone reference. In
Sec.~\ref{refembeddingLarge}, the asymptotic version of such
embeddings is studied in the large-sphere limit at null infinity.

\subsection{The definition of the Liu-Yau quasi-local energy}
\label{defLY}
\noindent
{\bf Definition 2.1}\\
\noindent
The {\it Liu-Yau quasi-local energy} associated with the 2-surface $S$
is 
$$
E(S)\equiv E^{\rm ref}(S)-E^{\rm phys}(S).
$$

\section{The small-sphere limit}
\label{small}
This section is devoted to gauging Liu-Yau quasi-local energy $E(S)$
against criterion {\bf C3}-(c) when $S$ is a small sphere, as defined
below. It is shown that $E(S)$ satisfies {\bf C3}-(c)-(i) for
non-vacuum, but deviates, in vacuum, from the expected value in
{\bf C3}-(c)-(ii) by an extra term of which the physical nature is yet
to be explored.

\subsection{Construction of the small sphere}
\label{constructionSmall}
The small sphere around an arbitrary point in the space-time is a
space-like level set of the null cone emanating from that point.
$\forall p \in M$, $\exists$ a normal neighborhood $U$ of $p$ in $M$
which uniquely determines a star-shaped neighborhood $\widetilde U$ of
$0$ in $T_pM$, such that the exponential map ${\rm exp}_p$ is a
diffeomorphism of $\widetilde U$ onto $U$ whose inverse is denoted by
${\rm exp}_p^{-1}$. Without loss of generality, it is assumed that
$\widetilde U$ is small enough such that the null cut locus
${\widetilde C}^+_N(p) \not\subset \widetilde U$ and hence that
$C^+_N(p)\not\subset U$. For a given orthonormal basis 
$\{e_i\}_{i=0}^{i=n-1}$ for $T_p M$ ($n=4$ when $M$ represents a
space-time), with $\langle e_i, e_j\rangle = \delta_{ij} 
\epsilon_j$, $(i, j = 0, \ldots, n-1)$ and its dual basis
$\{z^i\}_{i=0}^{i=n-1}$ for $T_p^*M$, the normal (Cartesian)
co-ordinate system of the connection $D$ on $U$ is defined by $\xi
\equiv (x^0, \ldots , x^{n-1}) \in {\bf
\mathfrak{F}}(U,{\mathds{R}}^n)$: $x^i=z^i \circ {\rm exp}^{-1}$,
$i= 0,\ldots, n-1$. Correspondingly, such a normal co-ordinate
system $(x^0, \ldots , x^{n-1})$ determined by
$\{e_i\}_{i=0}^{i=n-1}$ assigns to each point $q \in U$ co-ordinates
with respect to basis $\{e_i\}_{i=0}^{i=n-1}$ of the pull-back ${\rm
exp}_p^{-1}(q) \in \widetilde U \subset T_p M$ via ${\rm
exp}_p^{-1}(q)=\sum\limits_{i=0}^{n-1}x^i(q)e_i$, $\forall q \in U$.
Thus, the normal (Cartesian) co-ordinate system $(x^0, \ldots
,x^{n-1})$ on $U$ induces a Cartesian co-ordinate basis
$\Big\{\frac{\partial}{\partial
  x^i}\Big\}_{i=0}^{i=n-1} \in {\mathscr{X}}(U)$.

Define the Lorentz radius function
$\delta \in {\bf \mathfrak{F}}(U, {\mathds{R}})$ on $M$ at
$p$ as $\delta(q) \equiv |{\rm exp}_{p}^{-1}(q)|$, $\forall q \in
U$ and consider the geodesic ball
$\delta^{-1}(c) = \Big\{q \in U: \sum\limits_{i=0}^{n-1} \epsilon_i
\big(x^i(q)\big)^2 = c^2\Big\} \subset U$ in normal co-ordinates for
sufficiently small $|c| \geq 0$ and a hypersurface
$(x^0)^{-1}(t)$ for a given $t \in {\mathds{R}}$.
Then
$S(c,t) \equiv \delta^{-1}(c) \cap
(x^0)^{-1}(t)=r^{-1}(\sqrt{c^2-\epsilon_0 t^2})$
is a closed sub-manifold of $M$, where
$r \in {\bf \mathfrak{F}}(U, [0, +\infty))$ by
$r \equiv \sqrt{c^2-\epsilon_0 (x^0)^2}$
is the radial co-ordinate of the spherical normal co-ordinates.
In particular, for $c=0$, $\delta^{-1}(0)=\dot{J}^+(p, U)$
and thus
$\forall q \in U$, $\exists$ null
$X_1={\rm exp}_p^{-1}(q) \in \widetilde U$,
such that the ray $\rho : [0, a_0) \longrightarrow {\widetilde U}$ by
$\rho(x^0)\equiv x^0 X_1$, $x^0 \in [0, a_0)$, where
$[0, a_0)$ is the maximal domain of $\rho$,
uniquely determines a radial null geodesic from $p$ to $q$,
$\gamma^{X_1}: [0, a_0) \longrightarrow U$ by $\gamma^{X_1}={\rm
  exp}_p \circ \rho$. The superscript $X_1$ stresses that $\gamma$ is
the local flow of $X_1$, which is understood hereafter and the
superscript will most often be suppressed when confusion is
unlikely. It is noted in passing that as $c=0$, $x^0=r$, hence
$\gamma$ is well affinely parameterized by $r \in [0, a_0)$. The
local null-cone $\Lambda(p) \equiv \delta^{-1}(0)-p$ is then
foliated by
$\mathscr{F}_{\Lambda}=\Big\{\gamma |_{(0, a_0)}:
\lim\limits_{r\to 0}{\dot \gamma}(r)={\rm exp}_p^{-1}(q),
q \in \Lambda(p)\Big\}$.

\noindent
{\bf Definition 3.1} \\
\noindent For every given $t$, $S(c,t)=S(0,r)\cong S^2$, as a
regularly embedded space-like sub-manifold of $M$, is the desired
{\it small sphere}. For economy of notation, $S(0,r)$ will almost
always be abbreviated as $S(r)$.

\subsection{Moving frames on the small sphere}
\label{movingframesSmall}
The regular embedding of $S(r)$ into $M$ suggests an envisaged choice
of adapted orthonormal null frames on $U \cap \Lambda(p)$ as in
Sec.~\ref{def}.
For every given $r=x^0 \in (0,a_0)$,
$\forall q \in S(r) \hookrightarrow M$,
$\exists$ null frame
$\{X_i(q)\}_{i=1}^{i=n} \in {\mathds C}W_q$, where
$W_q=T_q M \cap T_q M^{\perp}$, with its dual
$\{\theta^i(q)\}_{i=1}^{i=n}\in {\mathds C}W_q^\ast$ adapted to $S$,
such that
${\mathds C}T_q M={\mathds C}T_qS(r)\oplus {\mathds C}T_qS(r)^{\bot}$,
where
${\mathds C}T_q S(r)={\rm span}_{\mathds{C}}\{X_3(q), X_4(q)\}$ and
${\mathds C}T_q S(r)^{\bot}=
{\rm span}_{\mathds{C}}\{X_1(q),X_2(q)\}$.
The principal bundle of null frames ${\bf F}(U\cap\Lambda(p))$
thus possesses a reduced structure group
$G={\mathds{C}}^\times \cong {\rm GL}(1, {\mathds{C}})
\subset {\rm SL}(2,{\mathds{C}})$, where the right group action
${\bf F}(U \cap \Lambda(p)) \times G
\longrightarrow {\bf F}(U \cap \Lambda(p))$ is given by
$u \centerdot z=(X^\prime_1,\ldots, X^\prime_4)=(X_1,\ldots,X_4)A(z)$,
where
$u=(X_1,\ldots,X_4)\in {\mathds C}W_q$,
$\forall q \in U \cap \Lambda(p)$, $z \in G$, and
$A(z)={\rm diag}(|z|^2, \frac{1}{|z|^2},
\frac{z}{\overline z}, \frac{\overline z}{z})
\in {\rm GL}(1,{\mathds{C}})\cong G$. Such construction of adapted
null frames is known, in physics literature, as the
Geroch-Held-Penrose (GHP) formalism \cite{GHP1973},
\cite{Ehlers1974}. 

A few geometrical properties follow almost transparently from the
preceding construction. For pedagogical purposes, however, it is
considered helpful to first recall some facts about degenerate
sub-manifolds of semi-Riemannian manifolds \cite{Kupeli1987}, which
will be preliminary for Sec.~\ref{large} as well.

Denoted by ${\mathds A}K$ the degenerate bundle over a degenerate
sub-manifold $(H, g_{\rm H})$ of a semi-Riemannian manifold $(M, g)$
equipped with a Levi-Civita connection $D$. It is known
that
${\mathds A}K={\mathds A}TH \cap {\mathds A}TH^{\perp}$ and  
${\mathds A}K^{\perp}={\mathds A}TH + {\mathds A}TH^{\perp}$. In
particular, when $(H, g_{\rm H})$ is a null sub-manifold of a
Lorentzian manifold $(M, g)$, ${\mathds A}K$ is the unique null
${\mathds A}$-line sub-bundle of ${\mathds A}TH$. Furthermore, if
$H$ is a null hypersurface, then 
${\mathds A}K={\mathds A}TH^{\perp}$.

\noindent
{\bf Definition 3.2}\\
\noindent 
$(H, g_{\rm H})$ is {\it irrotational} if $DU \in {\rm
End}_{\mathds A}({\mathds A}TH)$, $\forall U\in \Gamma({\mathds
A}K)$, where the {\it bundle homomorphism} $DU \in {\rm Hom}_{\mathds
A}({\mathds A}TH, {\mathds A}TM|_{\rm H})$, for a given $U \in
\Gamma({\mathds A}K)$, is defined as 
$DU(X) \equiv D_X U$, $\forall X \in {\mathds A}TH$.

The following properties are immediate consequences of the above
series of definitions.

\noindent
{\bf Proposition 3.3}\\
\noindent
(1) If $(H, g_{\rm H})$ is irrotational, then
$DU \in {\rm End}_{\mathds A}({\mathds A}TH)$ is self-adjoint,
$\forall U \in \Gamma({\mathds A}K)$, i.e.,
$g_{\rm H}(DU(X), Y)=g_{\rm H}(X, DU(Y))$,
$\forall X, Y \in {\mathds A}TH$. \\
\noindent
(2) If $(H, g_{\rm H})$ is integrable and
${\mathds A}K={\mathds A}TH^{\perp}$ (or, equivalently,
${\mathds A}K^{\perp}={\mathds A}TH$), then $(H, g_{\rm H})$ is
irrotational.

\noindent
Proof: \\
\noindent
(1) A straightforward calculation.\\
\noindent
(2) follows from (1). $\blacksquare$

\noindent
{\bf Corollary 3.4}\\
\noindent
Every degenerate hypersurface $(H, g_{\rm H})$ is
irrotational. $\blacksquare$ 

\noindent
{\bf Definition 3.5}\\
\noindent $U\in \Gamma({\mathds A}K)$ is {\it pregeodesic} if
$\exists f \in {\bf \mathfrak F}(H, {\mathds R})$, 
such that $D_U U=fU$.
$(H, g_{\rm H})$ is {\it geodesic} if every 
$U \in \Gamma({\mathds A}K)$ is pregeodesic. 

\noindent
{\bf Corollary 3.6}\\
\noindent
If $(H, g_{\rm H})$ is irrotational, then it is geodesic.

\noindent
Proof: compatibility and torsion-free properties of $D$. $\blacksquare$

In the present context, $\Lambda(p)$ is a degenerate (in fact,
null) hypersurface of $M$. Thus, with ${\mathds A}={\mathds C}$, the
following lemma, which will be used frequently throughout later
calculations, is readily seen to hold.

\noindent
{\bf Lemma 3.7}\\
\noindent
(1) $\varepsilon \equiv
\frac{1}{2}\Big(\theta^1(D_{X_1}X_1)-
i\theta^3(D_{X_1}X_3))\Big)
= 0$ and $\kappa \equiv \theta^2(D_{X_1}X_3) = 0$. \\
\noindent
(2) $\rho \equiv -\theta^2(D_{X_4}X_3)$ and
$\mu \equiv \theta^1(D_{X_3}X_4)$ are real. \\
\noindent
(3) $\tau-{\overline \alpha}-\beta =0$ and
$\pi-\alpha-{\overline \beta}=0$, where
$\tau \equiv -\theta^2(D_{X_2}X_3)$,
$\pi \equiv -\theta^2(D_{X_2}X_4)$
and
${\overline \alpha}+\beta \equiv -\theta^2(D_{X_3}X_2)$ \\
\noindent
Proof: \\
\noindent (1) Since $X_1={\dot \gamma}$ for the radial null geodesic
$\gamma$, $D_{X_1}X_1=0$. Hence the orthonormality of the null frame
implies, for the Levi-Civita connection $D$, that
$\theta^2(D_{X_1}X_2)=0$, $\theta^3(D_{X_1}X_3)=0$, and thus
the claim. \\
\noindent
(2) Since $\Lambda(p)$ is a null hypersurface of $M$, it follows
immediately from Proposition 3.3 that
$\theta^3(DX_1(X_3))=\theta^4(DX_1(X_4))$.\\
\noindent
(3) Similar to (2). $\blacksquare$

To better demonstrate the geometry of the local null-cone and the small
sphere, it is instructive to introduce angular co-ordinate functions
$(\vartheta, \varphi) \in {\bf \mathfrak{F}}
(U \cap
(\Lambda(p)-\{\gamma_0, \gamma_\pi\}), (0, \pi) \times [0,2\pi))$,
where $\gamma_0$ and $\gamma_\pi \in \mathscr{F}_{\Lambda}$
are two disconnected leaves (aka generators) of $\mathscr{F}_{\Lambda}$.
Together with $x^0$ and $r$, the local null-cone
$(\Lambda(p)-\{\gamma_0, \gamma_\pi\})$ comes equipped with spherical
normal co-ordinates $\varsigma \equiv (x^0, r, \vartheta, \varphi)$.
Gauss' lemma then guarantees that
 $\Big\{\frac{\partial}{\partial x^0},\frac{\partial}{\partial r},
\frac{\partial}{\partial \vartheta}, \frac{\partial}{\partial
  \varphi}\Big\} \in
{\mathscr{X}}(U \cap (\Lambda(p)-\{\gamma_0, \gamma_\pi\}))$
constitutes a set of mutually orthogonal spherical co-ordinate
basis. In particular, for every fixed $r \in (0, a_0)$, the small
sphere $S(r)$ can be parameterized by $(\vartheta, \varphi)$ or, as
$S^2\cong {\mathds{C}}P^1$, by the stereographic co-ordinates
$(\zeta, {\overline \zeta})$, with the understanding that the poles
are not covered.

In the spherical normal co-ordinates $\varsigma$, the null orthonormal
frame at every point has co-ordinate representation:
$X_i(q)=\sum\limits_{a=1}^{n}
X_i^a(q) \frac{\partial}{\partial \varsigma^a}\Big|_q$,
$i=1,\ldots,n$,
$\forall q \in U \cap (\Lambda(p)-\{\gamma_0, \gamma_\pi\}))$.
One noticeable simplification in the co-ordinate representation of
$X_1(q)$ is furnished by the foliation $\mathscr{F}_{\Lambda}$.
Recall from Sec.~\ref{constructionSmall} that
$\forall \gamma \in \mathscr{F}_{\Lambda}$, ${\dot \gamma}=X_1$,
and that the affine parameter of $\gamma$ can be arranged to be one of
the spherical normal co-ordinates, namely $\varsigma^r=r$. Therefore,
along the local flow, $\gamma^{X_1}$, of $X_1$ in $U$, it is clear by
definition that
$1=D_{X_1}r=X_1 r$. Then, $X_1^a=\delta^a_r$ in the
co-ordinate representation of $X_1$, and thus
$X_1=1\cdot \frac{\partial}{\partial r}$.

\subsection{Remarks on the vertex of the null cone}
\label{vertex}
Notice, by definition of normal co-ordinates, that
$x^0(p)=r(p)=0$, whereas the angular part of $\varsigma$ becomes
degenerate at $p$. Consequently, the moving frames cannot be extended
even continuously to $p$.
It is then necessary to examine the directional dependence of the
limiting process as $r \rightarrow 0$.

\noindent
{\bf Definition 3.8} \\
\noindent The {\it blow-up} \cite{GH1978} 
${\widehat \Lambda}(p)$ of $\Lambda(p)$ at
$p$ is defined as ${\widehat \Lambda}(p) \equiv \Lambda(p) \cup_\pi
S^2$ with the contraction $\pi: S^2 \longrightarrow p$. Hence
${\widehat \Lambda}(p)$ and $\delta^{-1}(0)$ are homotopy
equivalent. Correspondingly, the {\it blow-up} ${\widehat
{\widetilde U}_0}$ of ${\widetilde U}$ at $0$ is simply the
diffeomorphic pull-back of ${\widehat \Lambda}(p)$ by ${\rm
exp}_p^{-1}$, i.e., ${\widehat {\widetilde U}_0}=({\widetilde U}-0)
\cup_{\tilde \pi} {\widetilde S^2}$, where $\widetilde \pi:
{\widetilde S^2} \longrightarrow 0$ is the corresponding contraction
in $T_p M$.

Now, for a given orthonormal basis
$\{e_i\}_{i=0}^{i=n-1}$ of $T_p M$, it is natural to construct the
spherical radial basis vector
$e_r(\widetilde\vartheta, \widetilde\varphi) = {\rm
sin}{\widetilde\vartheta} {\rm cos}{\widetilde\varphi} e_1 + {\rm
sin}{\widetilde\vartheta} {\rm sin}{\widetilde\varphi} e_2 + {\rm
cos}{\widetilde\vartheta} e_3$, where $({\widetilde\vartheta},
{\widetilde\varphi}) \in (0, \pi) \times [0, 2 \pi)$ represents the
corresponding angular co-ordinates of ${\widetilde S^2}$ obtained
via the dual basis $\{z^i\}_{i=0}^{i=n-1}$ of $T_p^* M$. Then the
direction-dependent limit of the null basis
$\{X_i\}_{i=1}^{i=n} \in \Gamma({\mathds C}K)$
at ${\widetilde S^2}$ coincides with
$X_1(p,\widetilde\vartheta, \widetilde\varphi) = e_0+e_r$,
$X_2(p,\widetilde\vartheta, \widetilde\varphi) = \frac{1}{2}(e_0-e_r)$,
$X_3(p) =\frac{1}{\sqrt{2}}(e_1+i e_2)$, and $X_4 = {\overline X_3}$,
$\forall ({\widetilde\vartheta}, {\widetilde\varphi})
\in {\widetilde S^2}$.

Physically, the blow-up of $p$ defines an equivalence class of
spatially isotropic instantaneous observers at $p$. It suffices to
consider only linear isometries of $T_p M$, which constitute a
subgroup, ${\mathscr L}={\rm O}(3,1, {\mathds R})$, of the full
isometry group ${\rm I}(T_pM)={\mathscr L} \rtimes {\mathds R}^n$.

\noindent
{\bf Definition 3.9} \cite{SachsWu1977} \\
An {\it instantaneous observer} at $p$ is an ordered pair $(p,
e_0)$, where $e_0 \in T_p M$ is unit time-like and future-directed.
$(p,e_0)$ is said to be {\it spatially isotropic} if the stabilizer
of $e_0$ in ${\mathscr L}$ is ${\mathscr L}_{e_0}=O(3,{\mathds R})$.

\noindent
{\bf Definition 3.10} \\
\noindent An {\it equivalence class}, denoted as ${\mathscr
O}_p(e_0)$, of spatially isotropic instantaneous observers at $p$ is
defined as the orbit of ${\mathscr L}_{e_0}$ in $T_p M$.

It will be seen in Sec.~\ref{limitSmall} that it is
${\mathscr O}_p(e_0)$ who measures the quasi-local energy in the small
sphere limit at $p$.

While ${\widetilde S^2}$ is a topological 2-sphere, whether or not it
can be realized in the small-sphere limit as a metric
2-sphere depends upon whether $p$ is a curvature singularity, i.e.,
whether $p \in M$. 
Analysis in the cases where $p$ is of ``elementary singularity''
\cite{NewmanPosadas} is miserably complicated as no explicit
Riemannian metric on the limiting sphere ${\widetilde S^2}$ is
available. Efforts have been channeled
towards appealing to series expansions for small asphericity in
${\widetilde S^2}$ \cite{NewmanPosadas}, but have not proven to be as
promising as expected. On the other hand, when the
curvature at $p$ is finite, it is always possible to choose
${\widetilde S^2}$ as a metric 2-sphere as the limiting small
sphere. Therefore, only the latter situation is considered in the
present work.

\subsection{Reference embedding}
\label{refembeddingSmall}
The reference part of the quasi-local energy associated with the
2-surface $S(r)$ is solely determined by the intrinsic properties of
$S(r)$ although it appears to be defined in an extrinsic manner as in
Sec.~\ref{defref}. The apparent inconsistency is reconciled by virtue
of the integrability conditions for $S(r)$, particularly, as a
sub-manifold of $M$. It turns out that for the current problem at
hand, the Gauss equation alone is sufficient. 

As described in Sec.~\ref{defref}, consider the embedding 
$\iota^\circ: (S(r), g^{\rm S}, D^{\rm S})\hookrightarrow
(M^\circ, g^\circ, D^\circ)$ of $S(r)$ of dimension $m=2$ into
Minkowski $(M^\circ, g^\circ)$ with flat connection $D^\circ$.
Now, $\forall p \in S(r)$, with the usual identification of
${\mathds A}T_p S(r)$ and ${\mathds A}T_pS(r)^{\bot}$ with sub-spaces
of ${\mathds A}T_{\iota^\circ p} M$,
${\mathds A}T_{\iota^\circ p} M={\mathds A}T_p S(r) \oplus
{\mathds A}T_pS(r)^{\bot}$. Here, it is assumed
that $T_pS(r)^{\bot}={\rm span}_{\mathds{R}}\{T, N\}$,
with orthonormal frame $\{T, N\}$ and associated co-frame
$\{\theta^T, \theta^N\}$, where $T$ is chosen to be time-like and $N$ 
space-like. As a side remark, the standard space-time orthonormal
frame field is used in $TM$ (${\mathds A}={\mathds R}$), whereas the
null frame field is used in 
${\mathds C}TM$ (${\mathds A}={\mathds C}$). The same remark applies
dually for the co-frames. The transformation between the frame 
$\{T, N\}$ and the null frame $\{X_1, X_2\}$ as in
Sec.~\ref{movingframesSmall}. is, by convention, fixed by 
$(T, N)=(X_1, X_2) \mathfrak{U}$, where 
$\mathfrak{U}= \left(\begin{array}{cc} \frac{1}{2} & \frac{1}{2} \\
1 & -1 \end{array} \right) \in SO(1,1, {\mathds{C}})$.
Together with any given orthonormal frame
$\{E_j\}_{j=1}^{j=m} \in {\mathds A} T_pS(r)$ and its co-frame
$\{\omega^j\}_{j=1}^{j=m} \in {\mathds A} T_p^{\ast} S(r)$,
established are an adapted orthonormal frame
$\{{\overline E}_j\}_{j=1}^{j=n} \in {\mathds A} T_{\iota^\circ p}M$
and its co-frame
$\{{\overline \omega}^j\}_{j=1}^{j=n} \in
{\mathds A}T_{\iota^\circ p}^{\ast} M$
satisfying
$${\overline E}_j=
\left\{\begin{array}{r@{\quad:\quad}l}
T & j=1 \\ N & j=2 \\ E_j & j=3, 4
\end{array} \right.$$ and
$${\overline \omega}^j=
\left\{\begin{array}{r@{\quad:\quad}l}
\theta^T & j=1 \\ \theta^N & j=2 \\ \omega^j & j=3, 4
\end{array} \right.$$
Alternatively, if null frames are used,
$${\overline E}_j=
\left\{\begin{array}{r@{\quad:\quad}l}
X_j & j=1, 2 \\ X_j & j=3, 4
\end{array} \right.$$ and
$${\overline \omega}^j=
\left\{\begin{array}{r@{\quad:\quad}l}
\theta^j & j=1, 2 \\ \theta^j & j=3, 4
\end{array} \right.$$
Connections $D^{\rm S}$ and $D^\circ$ are related by the shape
tensor $s^{\bot}$. By definition,
$\forall x_p, y_p \in T_p S(r)$, $D^\circ{x_p}y=D^{\rm
S}_{x_p}y+s^{\bot}(x_p, y_p)$, where $s^{\bot}(x_p, y_p)=II^T(x_p,
y_p)T+II^N(x_p, y_p)N$, with $II^T(x_p, y_p)=\theta^T(D^\circ_{x_p}
y)$ and $II^N(x_p, y_p)=\theta^N(D^\circ_{x_p} y)$, where $y$ is the
extension of $y_p$ in the neighborhood of $p$.

The fully contracted Gauss equation for the embedding
$(S(r), g^{\rm S}, D^{\rm S}) \hookrightarrow
(M^\circ, g^\circ, D^\circ)$ can be written as
\begin{eqnarray}
0=S_{S(r)} & + & {\overline \epsilon}_2
\Big[\sum\limits_{j=3}^{4}\sum\limits_{l=3}^{4}
II^N(E_l, E_j) II^N(\omega^l \omega^j) - \big(II^N\big)^2\Big]
\nonumber \\
& + & {\overline \epsilon}_1
\Big[\sum\limits_{j=3}^{4}\sum\limits_{l=3}^{4}
II^T(E_l, E_j) II^T(\omega^l \omega^j) - \big(II^T\big)^2\Big], \nonumber
\end{eqnarray}
where
 ${\overline \epsilon}_1=g(T, T)$, ${\overline \epsilon}_2=g(N, N)$,
and $II^r =\sum\limits_{l=3}^{4}II^r(\omega^l, E_l), (r=N, T)$, or,
for brevity,
\begin{equation}
{\overline \epsilon}_2 \Big[(II^N)^2 - II^N \cdot II^N\Big]+
{\overline \epsilon}_1 \Big[(II^T)^2 - II^T \cdot II^T\Big]-
S_{S(r)}=0.
\label{ContractedGauss}
\end{equation}
Further simplification of the integrability conditions comes from the
use of the so-called null-cone reference (aka light-cone reference).
The same procedure as in Sec.~\ref{constructionSmall}. applies for the
construction of $\Lambda^{\circ}(p)$ in the reference space-time
  $M^\circ$ only now, it is a bona fide null cone as $M^\circ$ is
  Minkowski.  One of the immediate benefits is that the leaves of
$\mathscr{F}_{\Lambda^{\circ}}$ are shear-free, i.e.,
$-\sigma^{\circ}=\theta^2(D^\circ_{X_3}X_3)=
II^N(X_3, X_3)+II^T(X_3, X_3)=0=
-{\overline \sigma}^{\circ}=\theta^2(D^\circ_{X_4}X_4)=
II^N(X_4, X_4)+II^T(X_4, X_4)$. Then,
Eq(\ref{ContractedGauss}) becomes
\begin{equation}
{\overline \epsilon}_2 \Big[(II^N)^2 -
2 II^N(X_3, X_4)II^N(X_4, X_3)\Big]+
{\overline \epsilon}_1 \Big[(II^T)^2 -
2 II^T(X_3, X_4)II^T(X_4, X_3)\Big]-
S_{S(r)}=0.
\label{shearfree}
\end{equation}
Lemma 3.7-(2), when applied to $\Lambda^\circ (p)$, yields
${\overline \rho^{\circ}}=\rho^{\circ}$ and
${\overline \mu^{\circ}}=\mu^{\circ}$. Hence,
$II^N(X_3, X_4)=\theta^N(D^\circ_{X_3} X_4)=
-\frac{1}{2}{\overline \rho^{\circ}}+\mu^{\circ}=
-\frac{1}{2}\rho^{\circ} + {\overline \mu^{\circ}}=II^N(X_4, X_3)$.
Similarly,
$II^T(X_3, X_4)=II^T(X_4, X_3)$.
In fact, finer results of $\rho^{\circ}$ and $\mu^{\circ}$ can be
obtained by basis transformation ${\mathfrak{U}}$:
$\rho^{\circ}=-\theta^2({\overline D}_{X_4}X_3)=
-II^T(X_4, X_3) + II^N(X_4, X_3)
=\frac{1}{2} \big(-II^T+II^N\big)$ and, by the same token,
$\mu^{\circ}=\frac{1}{4}\big(II^T+II^N\big)$.
Hence,
$\|H^{\circ}\|=\sqrt{8\rho^{\circ}\mu^{\circ}}
=\sqrt{(II^N)^2-(II^T)^2}$. On the other hand,
$2II^N(X_3, X_4)II^N(X_4, X_3)=\frac{1}{2}\big(II^N\big)^2$ and
$2II^T(X_3, X_4)II^T(X_4, X_3)=\frac{1}{2}\big(II^T\big)^2$.
Therefore, Eq(\ref{shearfree}) can be written as
\begin{eqnarray}
& & \big(II^N\big)^2 - \big(II^T\big)^2-2S_{S(r)}=0, \nonumber
\end{eqnarray}
which is the ultimate integrability condition needed here. It then
follows that $\|H^{\circ}\|=\sqrt{2 S_{S(r)}}$, intrinsically
determined by the (positive) sectional curvature of $S(r)$ as
desired.

\subsection{Limiting process}
\label{limitSmall}
As declared in Sec.~\ref{vertex}, it is always assumed that curvature
at $p$ is finite. Thus, the blow-up of $p$, $S^2$, is a metric
2-sphere. Since the connection at $p$ is flat, the leading term in
every quantity is always its value in Minkowski space-time. The
limiting process is then a straightforward but tedious radial
expansion along the leaves of ${\mathscr{F}}_{\Lambda}$ of the
quasi-local energy associated with the small sphere $S(r)$ having a
standard $S^2$ as the limiting sphere at $p$.
Such expansions have been carried out, up to various accuracy order,
within the Newman-Penrose (NP) formalism \cite{NP1962} and are well
documented in the existing literature. In what follows, expansions of
related quantities are quoted without proof as details can be found
in, for example, Refs. \cite{KTW1986}, \cite{Dougan1992},
\cite{BLY1999}.\footnote{It is worth pointing out that Eq.(B18) in
  \cite{BLY1999} is a misprint, although the following Eqs.(B19) and
  (B20), are nevertheless, correct.}

To differentiate among the behaviors of the quasi-local energy in
non-vacuum and vacuum cases, the small-sphere limit is taken
separately. \\

\noindent
{\bf 1. Non-vacuum}

It turns out that the expansions of the relevant variables are needed
only along the leaves of ${\mathscr F}_{\Lambda}$ in the small-sphere
limit. In the spherical normal co-ordinates $\varsigma$ on the local
null cone of $p$, the following most pertinent quantities are
expanded, in the Newman-Penrose formalism, in power series of the
radial co-ordinate $r$.
\begin{eqnarray}
S_{S(r)} & = &2 r^{-2}+S^{(0)}_{S(r)} + O(r) \nonumber \\
\rho & = & - r^{-1}+ \frac{1}{3} r \phi^0_{00} + O(r^2) \nonumber \\
\mu & = & - \frac{1}{2} r^{-1} + \nonumber \\
& & \frac{1}{2} r \Big[\psi^0_2
+ {\overline \psi}^2_0 + 2 \Lambda^0 + \frac{2}{3}\phi^0_{11}
- \frac{1}{3} \phi^0_{00}\Big] + O(r^2) \nonumber \\
\Omega & = & \Omega_0 r^2
\Big[1 - \frac{1}{3} r^2 \phi^0_{00} + O(r^3) \Big], \nonumber
\end{eqnarray}
where $\Omega_0$ is the volume form of a metric 2-sphere
$S_0$. Consequently, the quasi-local energy in non-vacuum can be
written as
\begin{eqnarray}
E & = & \frac{1}{8 \pi} \varint\limits_{S(r)}
(\sqrt{2 S_{S(r)}}-\sqrt{8 \rho \mu}) \Omega \nonumber \\
& = & \frac{r^3}{2} \varint \limits_{S_0}
\Big[\frac{1}{3}(\phi^0_{00}+\eth_0 \phi^0_{10} +
{\overline \eth}_0 {\overline \phi}^0_{10} -
2 \eth_0 {\overline \psi}^0_{10} -
2 {\overline \eth}_0 \psi^0_1)+ \nonumber \\
& & (\psi^2_0 + {\overline \psi}^2_0 + 2 \Lambda^0 +
\frac{2}{3} \phi^0_{11} -
\frac{1}{3}\phi^0_{00}) + \frac{1}{3} \phi^0_{00}\Big]
\frac{\Omega_0}{4 \pi} + O(r^4). \nonumber
\end{eqnarray}
Most of the terms in the integral vanish for one of two reasons
shown below. The validity of the following lemma is well known,
although it is sketched here with a slightly more formal
proof. \\ 

\noindent
{\bf Lemma 3.11}\\
\noindent
(a) $\varint\limits_{S_0} (\eth_0 f) \Omega_0 = 0$,
$\forall f \in {\mathscr{C}}^k(-1)$ and
$\varint\limits_{S_0} ({\overline \eth}_0 g) \Omega_0 = 0$,
$\forall g \in {\mathscr{C}}^k(+1)$, where
${\mathscr{C}}^k(s)$ is the sheaf of germs of spin weight $s$
${\mathds{C}}$-valued functions.

\noindent
(b) $\varint\limits_{S_0} {\rm Re}(\psi^0_2) \Omega_0=0$.

\noindent
Proof: \\
\noindent
(1) An observation made in
\cite{EastwoodTod1982} indicates that, on $S_0 \cong
{\mathds{C}}P^1$, the elliptic operator $\eth$ is merely $\partial$ in
disguise. Hence,
$\eth:{\mathscr{C}}^k(-1) \longrightarrow {\mathscr{C}}^k(0)$ is
essentially the same as
$\partial: {\mathscr{E}}^{0, 1} \longrightarrow {\mathscr{E}}^{1, 1}$,
where ${\mathscr{E}}^{p, q}$ is the sheaf of germs of
${\mathds{C}}$-valued forms of type $(p,q)$ \cite{Wells1980}. A
similar argument holds for the conjugate operators. Then it is more
transparent that the integral vanishes essentially by virtue of
Stokes' theorem. \\
\noindent (2) By definition, ${\rm
Re}(\psi_2)=-\frac{1}{2}\theta^1(W_{X_1X_2}X_1)$, where $W$ is the
Weyl tensor. Then, using the basis transformation ${\mathfrak{U}}$
at $p$, i.e., on ${\widetilde S}^2$, $g(W_{X_1X_2}X_1,
X_2)=\frac{1}{4} {\mathscr W}((e_0+e_r)\wedge (e_0-e_r),
(e_0+e_r)\wedge (e_0-e_r)) = {\mathscr W}(e_0\wedge e_r, e_0\wedge
e_r)= a^2 {\mathscr W}(e_0\wedge e_1, e_0\wedge e_1) + b^2 {\mathscr
W}(e_0\wedge e_2, e_0\wedge e_2) + c^2 {\mathscr W}(e_0\wedge e_3,
e_0\wedge e_3) + 2 a b {\mathscr W}(e_0\wedge e_1, e_0\wedge e_2) +
2 a c {\mathscr W}(e_0\wedge e_1, e_0\wedge e_3) + 2 b c {\mathscr
W}(e_0\wedge e_2, e_0\wedge e_3)$. Here, with a slight abuse of
notation, $W$ is also represented by ${\mathscr W} \in {\rm
SymBil}(\wedge^2 T_pM \times \wedge^2 T_p M, {\mathds{R}})$, such
that ${\rm Ric}({\mathscr W})=0$ \cite{GHL1987}. In terms of the
spherical harmonics $Y^m_l({\widetilde \vartheta}, {\widetilde
\varphi})$, $a=\sqrt{\frac{2 \pi}{3}} (Y^{-1}_1 - Y^1_1)$, $b=i
\sqrt{\frac{2 \pi}{3}} (Y^1_1 + Y^{-1}_1)$, and
$c=\sqrt{\frac{2 \pi}{3}} Y^0_1$, where $({\widetilde \vartheta},
{\widetilde \varphi}) \in {\widetilde S}^2$ and $\{e_i\}_{i=0}^{i=3}
\in T_p M$ are as given in Sec.~\ref{vertex}. Now it follows from
the orthogonality of spherical harmonics and the trace-free
character of the Weyl tensor that the integral vanishes identically.
$\blacksquare$

Now, it is straightforward to establish the following:
\begin{eqnarray}
E(S) & = & \frac{r^3}{2} \varint \limits_{S_0}
\Big[\frac{1}{3}\phi^0_{00}+2 \Lambda^0+\frac{2}{3} \phi^0_{11}\Big]
\frac{\Omega_0}{4 \pi} + O(r^4) \nonumber \\
& = & \frac{r^3}{2}
\Big[\frac{1}{3}{\rm Ric}(e_0, e_0)+\frac{1}{6}S\Big]+O(r^4) \nonumber \\
& = & \frac{4 \pi}{3} r^3 {\mathfrak{T}}(e_0, e_0)+O(r^4), \nonumber
\end{eqnarray}
where $\mathfrak T$ is the energy-stress tensor of matter. As
anticipated, in non-vacuum, the leading contribution of $E(S)$ in the
small sphere limit at $p$ comes from the energy of matter, observed by
${\mathscr O}_p(e_0)$ (c.f. Definition 3.10). This is
physically reasonable because any form of gravitational energy-momentum
that is quadratic in curvature enters at higher orders in $r$.

\noindent
{\bf 2. Vacuum}

In the vacuum case, higher order expansions are inevitably
necessary. It is, nevertheless, a straightforward calculation to
obtain the following expansions:

\begin{eqnarray}
S_{S(r)} & = & 2 r^{-2} + S^{(0)}_{S(r)} + r S^{(1)}_{S(r)} +
r^2 S^{(2)}_{S(r)} + O(r^3) \nonumber \\
\rho & = & - r^{-1} +
\frac{1}{45} r^3 \psi^0_0 {\overline \psi^0_0} + O(r^4) \nonumber \\
\mu & = & -\frac{1}{2} r^{-1} +
\frac{1}{2} r (\psi^2_0 + {\overline \psi}^0_2)
+ \frac{1}{3} r^2 (\psi^1_2 + {\overline \psi}^1_2) + \nonumber \\
& & r^3 \Big(\frac{1}{360} \psi^0_0 {\overline \psi}^0_0 -
\frac{1}{40} \eth_0 ({\overline \psi}^0_0 \psi^0_1 +
4 {\overline \psi}^2_1) - \nonumber \\
& & \frac{1}{40} {\overline\eth}_0 (\psi^0_0 {\overline \psi}^0_1 +
4 \psi^2_1) - \frac{1}{4} S^{(2)}_{S(r)} \Big) + O(r^4), \nonumber \\
\Omega & = & \Omega_0 r^2 \Big[1 -
\frac{1}{90}r^4 \psi^0_0 {\overline \psi}^0_0 + O(r^5)\Big], \nonumber
\end{eqnarray}
where
\begin{eqnarray}
S^{(0)}_{S(r)} & = & -\frac{4}{3}(\eth_0 {\overline \psi}^0_1 +
{\overline \eth}_0 \psi^0_1) \nonumber \\
S^{(1)}_{S(r)} & = & -\frac{5}{6}(\eth_0 {\overline \psi}^1_1 +
{\overline \eth}_0 \psi^1_1) \nonumber \\
S^{(2)}_{S(r)} & = & \frac{1}{45} \psi^0_0 {\overline \psi}^0_0 -
\frac{3}{5}\eth_0 {\overline \psi}^2_1 -
\frac{3}{5}{\overline \eth}_0 \psi^2_1 -
\frac{17}{90} \eth_0 ({\overline \psi}^0_0 \psi^0_1) -
\frac{17}{90} {\overline \eth}_0
(\psi^0_0{\overline \psi}^0_1). \nonumber
\end{eqnarray}
In light of Lemma 3.11, the quasi-local energy in vacuum becomes
\begin{eqnarray}
E(S) & = & \frac{1}{8 \pi} \varint\limits_{S(r)}
(\sqrt{2 S_{S(r)}}-\sqrt{8 \rho \mu}) \Omega \nonumber \\
& = & \frac{r^5}{72} \varint\limits_{S_0}
{\overline \psi}^0_0 \psi^0_0 \frac{\Omega_0}{4 \pi} -
\frac{3}{4}\frac{r^5}{2}\varint\limits_{S_0}
({\rm Re}\psi^0_2)^2 \frac{\Omega_0}{4 \pi} + O(r^6) \nonumber \\
& = & \frac{r^5}{90} {\mathfrak{B}}(e_0, e_0, e_0, e_0)-
\frac{3}{4}\frac{r^5}{2}\varint\limits_{S_0}
({\rm Re}\psi^0_2)^2 \frac{\Omega_0}{4 \pi} + O(r^6), \nonumber \\
& = & \frac{r^5}{90} {\mathfrak{B}}(e_0, e_0, e_0, e_0) -
\frac{r^5}{40} \Big({\mathfrak E}^2(e_1, e_1) + {\mathfrak
E}^2(e_1, e_2) + {\mathfrak E}^2(e_1, e_3) + {\mathfrak E}^2(e_2,
e_3) \nonumber \\
& & - {\mathfrak E}(e_2, e_2) {\mathfrak E}(e_3, e_3)\Big) + O(r^6),
\nonumber
\end{eqnarray}
where ${\mathfrak {B}}$ is the Bel-Robinson tensor \cite{Bel} and
${\mathfrak E}(e_i, e_j)=-g(W_{e_0 e_j}e_0, e_i)$, $i, j = 1, 2, 3$,
is the symmetric (aka electric) part of the Weyl tensor. In the
leading order, $O(r^5)$, the first term is the gravitational energy
measured by ${\mathscr O}_p(e_0)$, whereas the second term comes
from the power series expansion in $r$ of $\sqrt{2S_{\rm S(r)}}$ in
the integrand of $E(S)$. The physical interpretation of this
additional term remains unclear. 

\section{The large-sphere limit at null infinity}
\label{large}
For the sake of simplicity and also of highlighting the physics, the
discussion is restricted to perfectly isolated sources in an empty
(i.e., Ricci-flat) space-time. An attempt at generalization to
non-isolated gravitational sources is briefly mentioned in
Sec.~\ref{Cmetrics}.

\subsection{Construction of the large sphere at null infinity}
\label{constructionLarge}
Several definitions and elementary properties of null infinity are
collected here for the purpose of unifying terminology and notation.\\

\noindent
{\bf Definition 4.1}\\
\noindent
A $C^r$ $(r\geq 0)$
{\it asymptote of a differentiable manifold} $M$ is an ordered
triple $({\widetilde M},i,f_\Omega)$, where ${\widetilde M}$ is a
manifold with boundary $\partial {\widetilde M}$, $i:M
\hookrightarrow {\widetilde M}$ is embedding by inclusion, and
$f_\Omega:\partial {\widetilde M}\times [0, +\infty) \longrightarrow
  {\widetilde M}$, such that $f_\Omega(x,0)=x$,
$\forall x\in \partial {\widetilde M}$, is a {\it collar}
\cite{Hirsch1976}.

The existence of collars in the differentiable category is easily
shown \cite{Hirsch1976}. That boundaries of $C^0$ manifolds have
collars is, however, far from obvious but is proved to be true
\cite{Brown1962}. A collar $f_\Omega$ can essentially be
characterized, with recourse to the partition of unity, if necessary,
by $\Omega \in {\bf {\mathfrak{F}}}(U, [0, +\infty))$, having $0$ as
its  regular value so that $\Omega^{-1}(0)=\partial {\widetilde M}$,
where $U \subset {\widetilde M}$ is a neighborhood of 
$\partial {\widetilde M}$. Hence, a collar is denoted hereafter simply
by $\Omega$.

A richer structure of the asymptote becomes available when a
differentiable manifold $M$ possesses a semi-Riemannian metric $g$ and
a Levi-Civita connection $D$. In particular, in the Lorentzian
category, it is a space-time, denoted by an ordered triple
$(M, g, D)$. The collar $\Omega$ of the asymptote of a space-time
then becomes crucially intertwined with the geometric structure.

\noindent
{\bf Definition 4.2}\\
\noindent A $C^r$ $(r\geq 0)$ {\it asymptote of a space-time} $(M,
g, D)$ is a $C^r$ asymptote of $M$ with a Lorentzian metric
${\widetilde g}$ and a Levi-Civita connection ${\widetilde D}$
associated to ${\widetilde M}$.

There exist in the literature various ways of defining the
asymptotic structure of null infinity, most of which are essentially
equivalent. It is attempted here yet another formulation that might be
more appropriate in logic. Elaborate discussions of the topological
properties of simple space-times can be found in, for example,
\cite{Newman1989}, whereas more geometric properties are recorded in,
for example, \cite{Kupeli1987}. For brevity, only future null infinity is
  considered for the past null infinity can be treated dually.

\noindent
{\bf Definition 4.3}\\
\noindent The {\it null infinity} (${\mathscr{I}}$) of the space-time
$(M, g, D)$ is a (not necessarily connected) null sub-manifold of
$\partial {\widetilde M}$ that is orientable and time-orientable.

In fact, the time orientation of $M$, and thus of ${\widetilde M}$,
induces a compatible time orientation of ${\mathscr I}$ so that
${\mathscr I}= {\mathscr I}^+ \cup {\mathscr I}^-$, where
${\mathscr I}^{\pm}=
{\mathscr I} \cap I^{\pm}({\widetilde M}, {\widetilde M})=
{\mathscr I}-I^{\mp}({\widetilde M}, {\widetilde M})$ are,
respectively, future (${\mathscr I}^+$) and past (${\mathscr I}^-$)
null infinity. Note that each of the
${\mathscr I}^{\pm}={\mathscr I}-{\mathscr I}^{\mp}$ is relatively
clopen in ${\mathscr I}$ and thus is a connected component of
${\mathscr I}$.

In particular, ${\mathscr I}$ has a collar, which, with a slight abuse
of notation, is also denoted by $\Omega$, as null infinity is the only
piece of $\partial {\widetilde M}$ that is of interest here. Moreover,
the conformal properties of null structure in the Lorentzian category
can be exploited to provide remarkable convenience in the analysis of
the asymptotic structure of a space-time.

\noindent
{\bf Definition 4.4}\\
\noindent A space-time $(M, g, D)$ is {\it asymptotically empty and
flat at null infinity} if \\
\noindent
(1) It is conformally diffeomorphic to its asymptote at null infinity
with $i^{\ast} {\widetilde g}=\Omega^2 g$ and
$i^{\ast}{\widetilde d}\Omega \neq 0$, where ${\widetilde d}$ is the
exterior derivative in ${\widetilde M}$; \\
\noindent
(2) $\exists$ ${\mathscr{I}}_0=\{q \in {\mathscr{I}}: q {\rm \ is\
  strongly\ causal}\} \subset {\mathscr I}$, which is thus open in
${\widetilde M}$ and is relatively open in ${\mathscr I}$; \\
\noindent (3) $\Omega^{-2}{\rm Ric}$ admits a $C^r$ extension to
$\Omega^{-1}(0)$; \\
\noindent
(4) ${\mathscr I}_0 \approx S^2 \times I$, where
$I\subset {\mathds R}$ is a connected component of
${\mathds R}$. \\
\noindent
$(M, g, D)$ is said to be 
{\it asymptotically Minkowskian at null infinity} when
$I={\mathds{R}}$. 

Definition 4.4 implies that ${\mathscr I}_0$ can be causally separated
by a space-like surface $S \approx S^2 \subset {\mathscr I}_0$. Hence
there exists a tubular neighborhood ${\mathscr N}_2$ of $S$ in
${\mathscr I}_0$, i.e., there exists a line bundle
$(E, S, \pi_{\rm S})$, where $E=TS^{\perp}$, and a $C^r$ diffeomorphism
$\psi: E \rightarrow {\mathscr N}_2$, with $\psi(0_x)=x$,
$\forall x \in S$, such that the diagram
$$
\xymatrix{
E \ar[d]_{\pi_{\rm S}} \ar[rd]^{\psi} & \\
S & {\mathscr N}_2 \ar[l]^{\rm ret} \\}.
$$
commutes \cite{Hirsch1976}, where ${\rm ret}: {\mathscr N}_2
\longrightarrow S$ is a retraction. Then, ${\mathscr N}_2$ is
foliated by curves that intersect $S$ transversely. Furthermore,
since ${\mathscr I}_0$ is a null hypersurface in ${\widetilde M}$,
it is irrotational and thus geodesic by Corollary 3.4 and 3.6.
Hence, $\forall$ future-directed null curve ${\widetilde \lambda} :
{\mathscr E}_2 \longrightarrow {\mathscr I}_0$, which can be
parameterized to be a null geodesic, $\exists a \in {\mathscr E}_2$,
such that ${\widetilde \lambda} \pitchfork_{q} S$, where
$q={\widetilde \lambda}(a)$. On the other hand, as ${\dot
{\widetilde \lambda}}(a)$ is null, ${\dot {\widetilde \lambda}}(a)
\notin T_qS$, which is space-like. Thus, either $q\notin S$ or $q
\in S$ and ${\widetilde \lambda}_{\ast a} (T_a {\mathscr E}_2) + T_q
S=T_q {\mathscr I}_0$. For notational purposes, denote $X_2={\dot
{\widetilde \lambda}}$, then $X_2 \in \Gamma ({\mathds C}L_2)$,
where $L_2=T{\mathscr I}_0\cap T{\mathscr I}_0^{\perp}$ is the only
null line bundle over ${\mathscr I}_0$. In fact, with a little
hindsight, the normal bundle $E$ over $S$ might as well be chosen to
be $L_2$.  
 
Now, ${\mathscr N}_2$ can be co-ordinatized by
$\varrho_2=(u, \vartheta, \varphi)\in {\bf \mathfrak{F}}
({\mathscr N}_2, {\mathscr E}_2 \times (0, \pi)\times [0,2\pi))$,
with the usual understanding that the poles of $S^2$ are not
covered. Clearly $S^2$ is a non-empty locally acausal compact
connected topological 2-sub-manifold of ${\mathscr I}_0$, hence is
sometimes called a ``cut'' of ${\mathscr I}_0$. The set of all such
cuts is denoted by ${\mathfrak C}_{{\mathscr I}_0}$.

To define the large sphere, first recall that a subset $F \subset M$
is a causal (resp. chronological) future set in a space-time $M$ if
$J^+ (F, M) \subset F$ (resp. $I^+(F, M) \subset F$). For a given
compact subset $K \subset M$, $J^+(K, {\widetilde M})$ is a causal
future set of ${\widetilde M}$. Consider ${\widetilde A}(K) \equiv
{\dot J}^+ (K,{\widetilde M})$. It is shown in \cite{Newman1989}
that ${\widetilde A}(K) \neq \emptyset$ and that ${\widetilde A}(K)$
is a compact achronal embedded $C^0$ sub-manifold of ${\widetilde
M}$ with boundary $\partial {\widetilde A}(K)={\widetilde A}(K)\cap
{\mathscr I}_0^+ \neq \emptyset$. With the differentiable structure
prescribed above, ${\widetilde A}(K)$ is a closed null sub-manifold
in ${\widetilde M}$ of co-dimension $1$ with boundary $\partial
{\widetilde A}(K)$. Therefore, ${\widetilde A}(K)$ has a tubular
neighborhood ${\mathscr N}_1$ in ${\widetilde M}$. Similar to the
treatment of ${\mathscr N}_2$, ${\widetilde A}(K)$ is geodesic and
${\mathscr N}_1$ is foliated by null geodesics ${\widetilde \gamma}:
{\mathscr E}_1 \longrightarrow {\widetilde A}(K)$ such that
$X_1={\dot {\widetilde \gamma}} \in \Gamma({\mathds C}L_1)$, where
$L_1 = T{\widetilde A}(K) \cap T{\widetilde A}(K)^{\perp}$ is the
only null line bundle over ${\widetilde A}(K)$.

\noindent
{\bf Definition 4.5}\\
The {\it large sphere near future null infinity} is a compact
2-surface $K \subset M$, such that ${\widetilde A}$ is a neat
sub-manifold \cite{Hirsch1976} and that ${\widetilde A}(K)\cap
{\mathscr I}_0^+ \in {\mathfrak C}_{{\mathscr I}_0^+}$.

According to Definition 4.5, $\forall x\in \partial {\widetilde A}$,
$T_x {\widetilde A}(K) \not\subset T_x {\mathscr I}_0^+$, i.e.,
${\widetilde A}(K)$ is nowhere tangent to ${\mathscr I}^+$, or,
notationally, ${\widetilde A}(K) \pitchfork {\mathscr I}^+$. On the
other hand, ${\mathscr I}^+$ has a collar, $\Omega$ say, which
restricts to a collar, $\Omega_{{\widetilde A}(K)}$, on $\partial
{\widetilde A}(K)$ in ${\widetilde A}(K)$. Hence, $\exists$ a null
foliation of ${\mathscr N}_1$ each of whose generators, ${\widetilde
\gamma}$, is affinely parameterized by $\Omega_{{\widetilde A}(K)}$
so that $D_{X_1} \Omega_{{\widetilde A}(K)}=1$, where $X_1={\dot
{\widetilde \gamma}}$, and that $\pounds_{X_1}
\vartheta=\pounds_{X_1}\varphi=0$, where $\pounds_{X_1}$ is the Lie
derivative along the flow of $X_1$. It is then natural to adopt
${\widetilde \varrho}= (u, \Omega, \vartheta, \varphi) \in {\bf
{\mathfrak F}}({\mathscr I}_0^+, {\mathscr E}_2 \times [0, +\infty)
\times (0, \pi)\times [0, 2 \pi))$ as a co-ordinate chart on the
neighborhood of ${\mathscr I}^+_0$ in the asymptote. By the
definition of a neat sub-manifold, ${\widetilde A}(K)$ is covered by
the chart $({\widetilde \varrho}, {\widetilde U})$ of ${\widetilde
M}$ such that ${\widetilde A} \cap {\widetilde U}={\widetilde
\varrho}^{-1}(u=u_0)$, where $u_0 \in {\mathscr E}_2$. Consequently,
${\widetilde A}$ admits an adapted co-ordinate system ${\widetilde
\varrho}_1= (u_0, \Omega_{{\widetilde A}(K)}, \vartheta, \varphi)
\in {\bf {\mathfrak F}}({\widetilde A}, [0, +\infty) \times (0,
\pi)\times [0, 2 \pi))$.

It is equally convenient to carry out the analysis in the asymptote
using the local chart ${\widetilde \varrho}$, as done in, for example,
\cite{Ludwig1976}.
However, for the purposes of studying ``physical fields''
\cite{Geroch1977}, it is more often useful to work in the original
space-time $(M, g, D)$. The latter approach is taken in what follows.

Most asymptotic behaviors of the original space-time near null
infinity come almost for free because $(M, g, D)$ is conformally
diffeomorphic to its asymptote. For example, ${\mathscr N}_1 \cap M$
in $M$ is foliated instead by null geodesics (possibly after
reparameterization of null pregeodesics) $\gamma:(b, +\infty)
\longrightarrow M$ such that ${\widetilde \gamma=i \circ \gamma}$
\cite{BEE1996}. Let $r$ be the affine parameter of $\gamma$, then
$D_{X_1} r=1$. The local co-ordinate chart in $M$ on the
neighborhood of null infinity is simply $\varrho=(u, r, \vartheta,
\varphi)$, which is known, in physics literature, as Bondi-type
co-ordinates, although in the original Bondi co-ordinates
\cite{BBM1962}, $r$ is chosen to be a luminosity distance parameter as
opposed to an affine parameter as used here.

\subsection{Moving frames on the large sphere}
\label{movingframesLarge} The seemingly pedantic construction in
Sec.~\ref{constructionLarge} exhibits its advantages now when it
comes to setting up adapted moving frames on ${\widetilde A}$;
almost the same moving frames as used in
Sec.~\ref{movingframesSmall}. can be applied in parallel for the 
large sphere $K$. The only difference lies in the obvious fact that
rather than the blow-up sphere $S^2$, the limit sphere now is
$K_0={\widetilde A}\cap {\mathscr I}^{+}_{0}\approx S^2$.
Correspondingly, all quantities in the Newman-Penrose formalism are
expanded in powers of $\frac{1}{r}$.

\subsection{Reference embedding}
\label{refembeddingLarge}
Given a generic space-time $(M, g, D)$, it is generally unlikely
that a large sphere $K$ in the asymptotic null region of $M$ could be
embedded into a genuine null cone in Minkowski reference space-time
that is emanated from one single point. Thus the embedding scheme in
Sec.~\ref{refembeddingSmall} becomes inappropriate in the large-sphere
limit. However, it is possible to isometrically embed
${\widetilde A}$ into the asymptotic null region of the Minkowski
space-time $(M^{\circ}, g^\circ, D^\circ)$
such that the sectional curvature of $K$, when calculated via the
Gauss equation, is preserved regardless of which ambient manifold into
which $K$ is embedded. Next, recall from Sec.~\ref{constructionLarge}
that the leaves of the foliation of ${\mathscr N}_1$ form a null
congurence of $X_1$. It is assumed \cite{WuChenNester2005} that the
shear of such null congurence is the same at $K_0$ in both $M$ and
$M^{\circ}$. The same construction as in Sec.~\ref{constructionLarge}
produces a local co-ordinate chart 
$\varrho^{\circ}=(u, r^{\circ}, \vartheta, \varphi)$ of the Bondi-type
in $M^{\circ}$ on the neighborhood of ${\mathscr I}_0^+$ that differs
from $\varrho$ only by a possible reparameterization of the null
congruence, registered by $r^{\circ}$ in $\varrho^{\circ}$.

Asymptotic expansions of the Newman-Penrose variables in both $M$ and
$M^\circ$ can be performed at one stroke. In $M^\circ$,
$\psi^{\circ}_i=0$, $i=0, \ldots, 4$ for $M^\circ$ is flat. Some of
the most pertinent expansions are listed below:
\begin{eqnarray}
\rho & = & -\frac{1}{r}-
\frac{\sigma^0 {\overline \sigma}^0}{r^3}
-\frac{(\sigma^0 {\overline \sigma}^0)^2
-\frac{1}{6}(\sigma^0 {\overline \psi}^0_0 + c.c.)}{r^5}
+O(r^{-6}) \nonumber \\
\mu & = & -\frac{1}{2 r}
-\frac{\psi^0_2 + \sigma^0 {\dot {\overline \sigma}}^0
+\eth^2_0 {\overline \sigma}^0}{r^2} + O(r^{-3}) \nonumber \\
\Omega & = & \Omega_0 r^2
\Big(1-\frac{\sigma^0 {\overline \sigma}^0}{r^2}\Big)
+O(r^{-2}) \nonumber \\
\rho^{\circ} & = &-\frac{1}{r^{\circ}}-
\frac{\sigma^0 {\dot {\overline \sigma}}^0}{{r^{\circ}}^3}
+O({r^{\circ}}^{-5}) \nonumber \\
\mu^{\circ} & = & -\frac{1}{2 r^{\circ}}
-\frac{\eth^2_0 {\dot {\overline {\sigma}^\circ}^0}}{{r^{\circ}}^2}
+O({r^{\circ}}^{-3}), \nonumber
\end{eqnarray}
where quantities with superscripts or subscripts $0$ represent their
corresponding asymptotic values at $K_0$, which are not to be confused
with those with superscripts $\circ$ in the reference space-time
$(M^\circ, g^\circ, D^\circ)$.

Expansions of the full set of Newman-Penrose variables in both $M$ and
$M^{\circ}$, when applied to the Gauss equation, establish an equality
of the sectional curvature of $K$, calculated via two different
embeddings, from which the relation between $r$ and $r^{\circ}$ can be
read off: $r^{\circ}=r+
(\eth^2_0 {\overline \sigma^\circ}^0
+{\overline \eth}^2_0 {\sigma^\circ}^0
-\eth^2_0 {\overline \sigma}^0
-{\overline \eth}^2_0 {\sigma}^0)+O(r^{-1})$.
Now the assumption
${\sigma}^0|_{K_0}={\sigma^\circ}^0|_{K^\circ_0}$ leads to a much
simpler relation $r^\circ = r + O(r^{-1})$.

\subsection{Bondi-mass loss}
\label{EBS}
After the preparatory work from previous sections, the calculation of
the quasi-local energy in the large sphere limit now becomes
completely transparent:
\begin{eqnarray}
E(K) & = & \frac{1}{8\pi} \varint\limits_{K}
[\sqrt{8 \rho^{\circ} \mu^{\circ}}- \sqrt{8 \rho\mu}] \Omega \nonumber \\
& = & -\frac{1}{4 \pi}\varint\limits_{K_0}
(\psi^0_2 +\sigma^0 {\dot {\overline \sigma}}^0) \Omega_0
+O(r^{-1}) \nonumber \\
& = & E_{\rm BS} + O(r^{-1}), \nonumber
\end{eqnarray}
where $E_{\rm BS}=-\frac{1}{4 \pi}\varint\limits_{K_0}
(\psi^0_2 +\sigma^0 {\dot {\overline \sigma}}^0) \Omega_0$ is the
Bondi mass loss.

\subsection{Energy flux}
\label{FBS} The energy flux through $K$ is defined as the rate of
change in the quasi-local energy $E(K)$ in the time-like direction
characterized by $T$. Here, $T(q)$ is related to
$\{X_i(q)\}_{i=1}^{i=2}\in {\mathds C}T_q K^{\perp}$, $\forall q \in
K$ by the same basis transformation ${\mathfrak U}$ as in
Sec.~\ref{refembeddingSmall} and agrees with the generator of the
time translation subgroup of the Bondi-Metzner-Sachs (BMS) group at
${\mathscr I}^+$ \cite{Penrose1974}. Hence,
\begin{eqnarray}
\frac{\partial E^{\rm phys}(K)}{\partial t} & = &
\frac{1}{8\pi} \varint\limits_{K}
\pounds_T (\sqrt{8\rho\mu}\Omega) \nonumber \\
& = & \frac{1}{8\pi} \varint\limits_{K}
\Big[\sqrt{8\rho\mu} \iota^\ast \Big(\pounds_T\Omega \Big)+
(T \sqrt{8\rho\mu}) \Omega \Big] \nonumber \\
& = & \frac{1}{8\pi} \varint\limits_{K}
\Big[\sqrt{8\rho\mu}(2 \mu -\rho)+ \nonumber \\
& & \Big((X_2+\frac{1}{2} X_1)\sqrt{8\rho\mu}\Big)\Big]\Omega
\nonumber \\
& = & \frac{1}{8 \pi} \varint\limits_{K_0}
2 \frac{\partial}{\partial u}
(\psi^0_2 + \sigma^0 {\dot {\overline \sigma}}^0
+\eth^2_0 {\overline \sigma}^0) \Omega_0
+O(r^{-1}), \nonumber
\end{eqnarray}
in which $\iota: K \hookrightarrow M$ is the embedding map that
induces
$\iota^{\ast} (\pounds_{X_1} \theta^3 \wedge \theta^4)=
-2\rho \theta^3 \wedge \theta^4$ and
$\iota^{\ast} (\pounds_{X_2} \theta^3 \wedge \theta^4)=
2\mu \theta^3 \wedge \theta^4$ by straightforward calculations.

Very similarly,
\begin{eqnarray}
\frac{\partial E^{\rm ref}(K)}{\partial t} & = &
\frac{1}{8\pi} \varint\limits_{K}
\pounds_T (\sqrt{8\rho^{\circ} \mu^{\circ}}\Omega) \nonumber \\
& = & \frac{1}{8 \pi} \varint\limits_{K_0}
2 \frac{\partial}{\partial u}
\eth^2_0 {\overline \sigma^{\circ}}^0 \Omega_0
+O(r^{-1}), \nonumber
\end{eqnarray}
where, again, $r^\circ=r+O(r^{-1})$ is used in the co-ordinate
representation of $X_1$ and $X_2$.

Recall that the embedding scheme is tacitly chosen so that
${\sigma^{\circ}}^0|_{K_0}=\sigma^0|_{K_0}$ and that $\psi^0_3 \in
{\mathscr B}(-1)$. Then, with the help of the Newman-Penrose
equations ${\dot \psi}^0_2 + \frac{1}{\sqrt{2}}\eth \psi^0_3 -
\sigma^0 \psi^0_4=0$ and $\psi^0_4+ {\ddot {\overline \sigma}^0}=0$,
the flux of quasi-local energy in the large-sphere limit at
${\mathscr I}^+$ is
\begin{eqnarray}
\frac{\partial E(K)}{\partial t} & = &
\frac{\partial E^{\rm ref}(K)}{\partial t}-
\frac{\partial E^{\rm phys}(K)}{\partial t} \nonumber \\
& = & -\frac{1}{4 \pi}\varint\limits_{K_0}
{\dot \sigma^0} {\dot {\overline \sigma}^0} \Omega_0
+O(r^{-1}), \nonumber
\end{eqnarray}
in which the leading term is precisely the flux of Bondi news.

\subsection{Remarks on generalization to non-isolated systems}
\label{Cmetrics}
Applying the notion of quasi-local quantities, in general, to
non-isolated gravitational systems may incur failure to satisfy, for
example, criterion {\bf C3}-d. The obstruction to such generalizations
largely lies in the difficulty of having a well-behaved or,
well-described large-sphere limit. It suffices to analyze the problem
at null infinity; the situation at the spatial infinity, if treated
with care, is quite similar. Intuitively, it is conceivable that
the non-isolated source has to eventually run off the boundariless
manifold and wreck the topological structure of the asymptote
described in Sec.~\ref{constructionLarge}. However, the following
example illustrates a more subtle cause.

\noindent
{\bf Definition 4.6}\\
\noindent Given two quartic polynomials $G(x)=1-x^2-2mA x^3-e^2 A^2
x^4$ and $F(y)=-G(-y)$, where $m\geq 0$, $e\in {\mathds R}$, and
$A>0$, assume that all of the roots of $G(x)$ are distinct among
which at least two are real (i.e., $e \neq 0$ or
$mA<\frac{1}{\sqrt{27}}$), denoted by $x_2<x_1$, such that $G(x)
\geq 0, \forall x \in [x_2,x_1]$. The {\it (charged) C-metrics},
denoted by $C_1=C_1(m, A, e)$, are a
3-parameter family of space-times of Petrov type-D, satisfying \\
\noindent
(1) topologically, $C_1 \cong P_1 \times Q_1$,
where
$P_1 \cong {\mathds R} \times {\mathds R}^+$ and $Q_1 \cong S^2$; \\
\noindent
(2) metrically, $(C_1, g_1, D_1)$ is conformally diffeomorphic to its
asymptote. In a local chart $\upsilon_{{\rm A}_1}=(t,y,x,z)$, set
$\kappa$ to be a real constant,
${\widetilde C}_1 \approx {\widetilde P}_1 \times {\widetilde Q}_1$,
where $({\widetilde P}_1, {\widetilde f}_1)$ is given by
${\widetilde P}_1 \approx {\mathds R}\times [-x,+\infty)$ and
${\widetilde f}_1=-F(y) dt\otimes dt + \frac{dy \otimes dy}{F(y)}$,
$({\widetilde Q}_1, {\widetilde h}_1)$ is given by
${\widetilde Q}_1 \approx (x_2, x_1) \times [0, 2\pi \kappa)$ and
${\widetilde h}_1=\frac{dx \otimes dx}{G(x)}+G(x) dz\otimes dz$.
Hence,
${\widetilde g}_1={\widetilde f}_1+{\widetilde h}_1$. The collar is
defined by $\Omega^{-1}=A(x+y) \in {\mathds R}^+$.

The following lemma exposes one of the most peculiar features of the
C-metrics, namely the so-called ``conical singularity'' or 
``nodal singularity'' \cite{KinnersleyWalker1970},
\cite{AshtekarDray1981} at one of the boundaries of the annulus
${\widetilde Q}_1$ (in the proof, $x_1$) that cannot be compactified
in the $C^r$ category for $r>0$. \\  
\noindent
{\bf Lemma 4.7}\\
\noindent
${\widetilde Q}_1 \cong S^2$ only in the $C^0$ category but
${\widetilde Q}_1 \cong D^2$ in the $C^r$ ($r>0$) category, \\
\noindent
Proof: \\
Consider another local chart
$\upsilon_{{\rm B}_1}=(t,y,\vartheta,\varphi)$
which is $C^r$-compatible with $\upsilon_{{\rm A}_1}$, where
$r \geq 0$, $\vartheta(x)=
\varint\limits^{x}_{x_2}\frac{dx^\prime}{\sqrt{G(x^\prime)}}$ and
$\varphi=\kappa^{-1} z$. Put
$\rho(\vartheta)=\sqrt{G(x(\vartheta))}$ and
$\vartheta_0=\vartheta(x_1)$. Then the graph of
$(\vartheta, \rho(\vartheta))$ is
$(0, \vartheta_0)\times (0,\rho_{\rm M}]$,
for some $\rho_{\rm M}>0$.
Clearly, $\rho (0)=\rho (\vartheta_0)=0$.
Now ${\widetilde Q}_1 \approx (0, \vartheta_0) \times [0, 2 \pi)$. \\
\noindent
(1) $r=0$, simply extend continuously the domain of
$\vartheta$ to compactify the annulus ${\widetilde Q}_1$. \\
\noindent (2) $r>0$, the standard Bertrand-Puiseux test
\cite{SpivakII}, when applied on ${\widetilde Q}_1$, shows that
\cite{KinnersleyWalker1970} the $C^r$-differentiable structure is
preserved at the boundary of the annulus ${\widetilde Q}_1$ if and
only if $\kappa^{-1}=\Big|\frac{d\rho}{d\vartheta}\Big|_{0,
\vartheta_0}$. However, $\Big|\frac{\partial \rho}{\partial
\vartheta}\Big|_{\vartheta_0} \neq \Big|\frac{\partial
\rho}{\partial \vartheta}\Big|_0$ unless $m=0$ or
$|e|=m>\frac{1}{4A}$. Thus, without loss of generality, set
$\kappa^{-1}=\Big|\frac{d\rho}{d\vartheta}\Big|_{\vartheta=0}$.
Then, except for two 2-parameter families of electrovac solutions,
$m=0$ or $|e|=m>\frac{1}{4 A}$, only one of the boundaries of the
annulus can be compactified such that 
${\widetilde Q}_1 \cong [x_2, x_1)\times [0,2 \pi) \cong D^2$. 
$\blacksquare$

Lemma 4.7 clearly shows that a generic C-metric is not asymptotically
empty and flat at null infinity in the sense of Definition 4.4. The
remedy comes out of a key observation \cite{AshtekarDray1981} that
$C_1$ is not maximal. Hence, a maximal extension of $C_1$ leads to
$C=C_1 \cup C_2$, where $C_2$ is an identical replicate of $C_1$. In
the asymptote of $C_2$,
${\widetilde C}_2 \approx {\widetilde P}_2 \times {\widetilde Q}_2$,
where Lemma 4.7 applies except that in the local chart
$\upsilon_{{\rm B}_2}$ of ${\widetilde Q}_2$ the opposite boundary of the
annulus ${\widetilde Q}_2$ is compactified, i.e.,
${\widetilde Q}_2 \cong (x_2, x_1]\times [0,2 \pi) \cong D^2$ in the
$C^r$ category for $r>0$. Therefore, as shown in
\cite{AshtekarDray1981}, in $C$, or its asymptote ${\widetilde C}$,
$\upsilon_{\rm B}=\{\upsilon_{{\rm B}_1}, \upsilon_{{\rm B}_2}\}$
constitutes an atlas for
${\widetilde Q}={\widetilde Q}_1 \cup {\widetilde Q}_2 \cong S^2$ and
that ${\mathscr I} \approx S^2 \times I$, where $I \cong {\mathds R}$
except for two null generators corresponding to precisely the
boundaries of the annulus ${\widetilde Q}$. But these two generators
can be arranged such that they both are incomplete in the future or in
the past.

The above analysis demonstrates that for a non-isolated source modeled
by the C-metrics, the topological and geometrical properties of
${\mathscr I}$ as in Definition 4.4 can certainly be retained almost
as for the isolated sources. The real difficulty, nonetheless, arises
at a technical level when practical calculations are carried
out. Evaluations of the Bondi mass or energy flux inevitably involve
integrations on ${\widetilde Q} \in {\bf {\mathfrak C}}_{\mathscr I}$,
which has to be covered by two charts. This task usually turns out to
be analytically intractable \cite{DrayThesis}.
 
\section{Summary and discussion}
\label{summary} The investigation of the limiting behavior of the
Liu-Yau quasi-local energy is carried out. Such an analysis could be
utilized to provide an appropriate certification for the Liu-Yau's
proposal as a physically sound candidate for the quasi-local energy.
Preliminary results that are considered new include:

\begin{itemize}
\item
In the small-sphere limit, the leading term in the
quasi-local energy measured by the equivalence class of
spatially-isotropic instantaneous observers ${\mathscr O}_p(e_0)$ at
an arbitrary point $p$ in non-vacuum is found to be
$\frac{4 \pi}{3} r^3 {\mathfrak T}(e_0, e_0)$, where
${\mathfrak T}$ is the energy-stress tensor of matter and $r$ is the
radius of the small sphere in the limit $r \rightarrow 0$.
\item
In vacuum, however, the gravitational quasi-local energy measured by
${\mathscr O}_p$ gains an extra term in the leading order, in
addition to the currently known limit
$\frac{r^5}{90} {\mathfrak B}(e_0, e_0, e_0, e_0)$, where
${\mathfrak B}$ is the Bel-Robinson tensor.

The occurrence of the extra term is considered as an example
of how the quasi-local energy depends rather crucially upon the choice
of the reference embedding. Since the co-dimension 2 embedding of the
2-surface $S$ into the reference space-time $M^\circ$ is in general
non-unique, it is plausible that an embedding scheme other than
the null-cone reference may result in a different limiting
behavior. Moreover, the currently known limit 
$\frac{r^5}{90} {\mathfrak B}(e_0, e_0, e_0, e_0)$ in vacuum is
actually model dependent and usually variable to reference embedding
(for example, \cite{BLY1999}). Therefore, it is contemplated that the
mismatch in the small-sphere limit in vacuum does not seem to serve as
a strong piece of evidentiary support to rule out Liu-Yau's model.
\item
In the large-sphere limit at null infinity of an asymptotically
empty and flat space-time, the Liu-Yau quasi-local energy is found
to coincide, in radiating scenarios, with the Bondi mass loss and
the news flux. A tentative generalization of the quasi-local energy
to non-isolated sources encounters technical difficulties at null
infinity in the example of the C-metric.
\end{itemize}
\acknowledgments
I wish to thank Professor R. R. Caldwell for suggesting the project
and for his guidance throughout the course of the investigation. I am
especially indebted to Professors V. Chernov and D. L. Webb for
numerous invaluable discussions. In particular, I thank Professor
V. Chernov for suggesting the use of the blow-up technique in
Sec.~\ref{vertex}.  This work was supported in part by NSF AST-0349213.


\end{document}